\documentclass[aps,pra,twocolumn,showpacs,preprintnumbers,amsmath,amssymb,superscriptaddress,10pt]{revtex4-1}

\usepackage{colortbl}
\usepackage[table]{xcolor}
\usepackage{tabularx}
\usepackage{amsmath}
\usepackage[latin1]{inputenc}
\usepackage{bm}
\usepackage{graphicx}
\usepackage{SIunits}
\usepackage{subfigure}
\usepackage{enumerate}
\usepackage{multirow}
\usepackage{color} 

\definecolor{darkblue}{rgb}{0,0,0.5}
\definecolor{lila}{rgb}{0.3,0,0.3}
\definecolor{turq}{rgb}{0,0.1,0.4}
\usepackage[colorlinks=true,
linkcolor=darkblue, 
  filecolor=red,
  citecolor=turq, 
  urlcolor=lila 
]{hyperref} 

\newcommand{\be}{\begin{equation}}
\newcommand{\ee}{\end{equation}}
\newcommand{\ba}{\begin{eqnarray}}
\newcommand{\ea}{\end{eqnarray}}
\newcommand{\ban}{\begin{eqnarray*}}
\newcommand{\ean}{\end{eqnarray*}}

\begin{document}

\title{A universal setup for active control of a single-photon detector}

\author{Qin Liu}
\affiliation{Department of Electronics and Telecommunications, Norwegian University of Science and Technology, NO-7491 Trondheim, Norway}
\author{Ant{\'i}a Lamas-Linares}
\affiliation{Centre for Quantum Technologies and Department of Physics, National University of Singapore, 3 Science Drive 2, Singapore 117543, Singapore}
\author{Christian Kurtsiefer}
\affiliation{Centre for Quantum Technologies and Department of Physics, National University of Singapore, 3 Science Drive 2, Singapore 117543, Singapore}
\author{Johannes Skaar}
\affiliation{Department of Electronics and Telecommunications, Norwegian University of Science and Technology, NO-7491 Trondheim, Norway}
\author{Vadim Makarov}
\email{makarov@vad1.com}
\affiliation{Institute for Quantum Computing and Department of Physics and Astronomy, University of Waterloo, 200 University Avenue West, Waterloo, Ontario N2L~3G1, Canada}
\author{Ilja Gerhardt}
\email{ilja@quantumlah.org}
\affiliation{Max Planck Institute for Solid State Research, Heisenbergstra\ss e 1, D-70569 Stuttgart, Germany}

\date{\today}

\begin{abstract}
The influence of bright light on a single-photon detector has been described in a number of recent publications. The impact on quantum key distribution (QKD) is important, and several hacking experiments have been tailored to fully control single-photon detectors. Special attention has been given to avoid introducing further errors into a QKD system. We describe the design and technical details of an apparatus which allows to attack a quantum-cryptographic connection. This device is capable of controlling free-space and fiber-based systems and of minimizing unwanted clicks in the system. With different control diagrams, we are able to achieve a different level of control. The control was initially targeted to the systems using BB84 protocol, with polarization encoding and basis switching using beamsplitters, but could be extended to other types of systems. We further outline how to characterize the quality of active control of single-photon detectors.
\end{abstract}

\maketitle

\section{Introduction}
The optical control of avalanche photodiodes (APDs) has been discussed recently, and was also implemented experimentally. It allows to control passively-~\cite{makarov_njp_2009}, actively-quenched~\cite{sauge_oe_2011} and gated~\cite{lydersen2010a,lydersen2010b} avalanche photodiodes. The electrical output of superconducting nanowire single-photon detectors (SNSPDs) can also be influenced~\cite{lydersen_njp_2011,tanner2013}. The high degree of control allows to intrude into quantum key distribution (QKD) setups~\cite{gerhardt_nc_2011} or to change the outcome of generic quantum optical experiments such as Bell tests up to non-physical values~\cite{gerhardt_prl_2011}. All experiments that rely on the detection of single photons can be influenced. The underlying theory has been outlined in a number of papers~\cite{makarov_njp_2009,sauge_oe_2011,lydersen2010a}: experiments require an electronic circuit which provides distinct control over a number of light sources.

QKD implementations rely often on encoding the qubits into polarization states of single photons. In this scenario, the avalanche photo detectors can be influenced by polarized bright laser pulses. To achieve a full level of control, it is required to target different detectors, which are combined into a complete polarization analyzer. For QKD experiments the so-called ``faked-state'' attack has been developed~\cite{makarov2005}. In this scheme, an eavesdropper Eve receives approximate single photons sent out by the legitimate sender (Alice), analyzes and saves the measurement outcome. Immediately after, Eve sends a tailored light pulse onwards to the legitimate receiver (Bob). Thereby, Eve has full knowledge on the legitimate connection of Alice and Bob. Since Bob's detector confuses the received pulse with a single-photon click, nothing indicates an intrusion into the key distribution scheme. Most security proofs do not cover this intrusion mechanism, since is acts on a classical part of single-photon detection. The latter are represented by electrical pulses, which we name ``clicks'' through the paper.

In this paper, we provide a detailed technical description of a universal detector control unit. This device was used to influence the outcome of several experiments~\cite{gerhardt_nc_2011,gerhardt_prl_2011}. Several ways to characterize the fidelity of control of a targeted system are described. The unit can be used to launch pulses into a variety of quantum-optical measurement setups. The system was designed to perform the entire optical and electronic control, and tp be compact and portable ($\approx 15\,\kilo\gram$). Therefore, the unit can be used at different setups on-site.

\section{Theory of operation}

To detect single photons, an APD is operated in Geiger mode, biased at a voltage slightly below the breakdown voltage with no illumination. An avalanche breakdown happens when an electron-hole pair, which is created by an absorbed photon, multiplies. During avalanche, macroscopic currents flows through the APD. When these currents exceeds a comparator threshold, an electrical pulse is produced at the detector output. Afterwards, the voltage across the APD is reduced below the breakdown voltage, to stop the avalanche~\cite{cova2004}. This so-called quenching can be introduced by various methods: active quenching reduces the bias voltage, passive quenching utilizes the finite recharge time of the device itself, and gating the APDs reduces the voltage periodically. During quenching, the APD is converted into a classical linear detector, i.e., the current through the APD depends linearly on the incident optical power~\cite{lydersen2010a}. After quenching, the voltage across the APD recovers back to the bias voltage, and the detector becomes ready for signaling the next photon.
s

\begin{figure}
  \includegraphics[width=70mm]{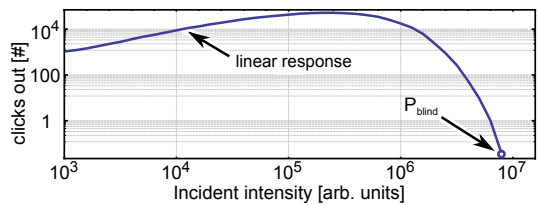}
  \caption{(Color online) Non-linear response from a single-photon detector.}
  \label{fig:detresp}
\end{figure}

In a typical single-photon detector, the click rate increases linearly up to saturation (at sometimes up to several million clicks per second) as the number of incident photons per second is increased. If the input number of photons is increased above saturation, the electrical response decays quickly. Although the amount of incident light is increased. The basic reason for this is that the APD is no longer sensitive to single photons as in the linear regime. Powerful illumination can suppress the APD's voltage below the breakdown voltage, since there is no full recovery between individual photons. After a certain threshold blinding power $P_{\rm blind}$, the detector falls completely silent, since the avalanches become too small to exceed the comparator threshold. For an experimental response function of a single photon counting avalanche photo diode, please refer to Fig.~\ref{fig:detresp}.

Several methods have been demonstrated to launch a click in such a ``blinded'' detector~\cite{makarov_njp_2009,sauge_oe_2011,lydersen2010a,lydersen2010b}. A simple method to launch a single click is to temporarily reduce the incident light power from above $P_{\rm blind}$ to zero~\cite{makarov_njp_2009}. The detector recovers some sensitivity, and interprets the subsequent rise of illumination back to above $P_{\rm blind}$ as a single-photon detection event. Another method, mostly used in this paper, is to constantly illuminate the detector by a power above $P_{\rm blind}$, and to apply a short much brighter pulse. This causes an additional photocurrent in the linear mode, which is sufficient to cross the comparator threshold and to produce a click~\cite{lydersen2010a,gerhardt_nc_2011}.

In addition to APDs, SNSPDs have also been used in modern QKD setups. SNSPDs can be controlled in a very similar way~\cite{lydersen_njp_2011,tanner2013}. Our detector control apparatus can be easily adjusted to be applied to a QKD setup based on SNSPDs.  

\begin{figure}
  \includegraphics[width=\columnwidth]{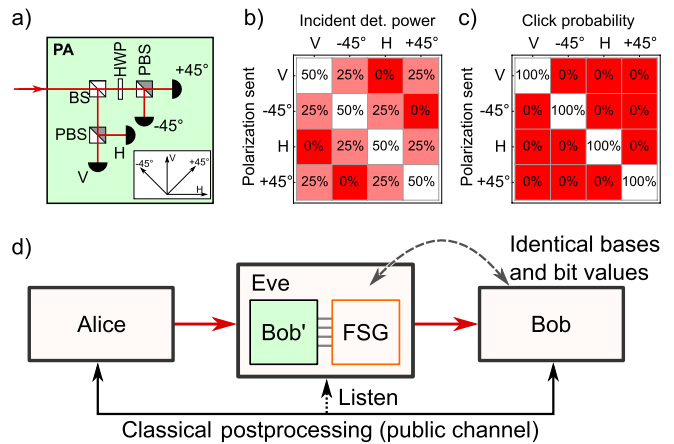}
  \caption{(Color online) QKD using polarization encoding, and faked-state attack on it. The qubits are encoded into polarization states of single photons (a, inset). Each of the four polarization states is detected by one single-photon detector. (a) The polarization analyzer can detect four different polarizations: horizontal (H), vertical (V), and similarly for the $45\degree$ tilted basis ($\pm 45\degree$). (b) When the detector receives a linear input polarization, the light is distributed unequally among the detectors. (c) Ideally, under attack the click probability should reach 100\% for each targeted detector and vanish for untargeted detectors. (d) The implemented attack uses the so-called faked-state scheme. In this scheme, the eavesdropper (Eve) detects the photon with a device Bob$'$ similar to the legitimate receiver (Bob), then sends a \emph{faked state} onwards to the latter. The faked state can generate the identity matrix (c) or any other matrix at Bob, at the eavesdropper's will. FSG, faked-state generator; BS, beamsplitter; PBS, polarizing beamsplitter; HWP, half-wave plate rotated at $22.5\degree$ angle.}
  \label{fig:polana}
\end{figure}

One question prior to the development of the control apparatus was whether a real detection apparatus, practically used in QKD, can be fully controlled. A typical detection apparatus, known as polarization analyzer~\cite{rarity1994}, is shown in Fig.~\ref{fig:polana}(a). It contains four single-photon detectors, each of which detects photons with one of the four polarization orientations~\cite{kurtsiefer_n_2002,kurtsiefer__2002} (see inset in Fig.~\ref{fig:polana}(a)). When a faked state is sent, at a certain detector, also the other three might click with different probabilities, since they might receive a certain amount of light from the control setup. The exact amount of incident power on each detector depends on the incoming polarization. To target only one detector, it is possible to send only a distinct linear polarization, such that only one detector sees 50\% of the incoming light. In the other basis, rotated by $\pm 45\degree$, each of the two detectors receives 25\% of the light. The fourth detector is orthogonally oriented to the incoming light, such that ideally, it does not receive any light (see Fig.~\ref{fig:polana}(b) for a distribution of powers).

The overall control efficiencies can be represented by a matrix of probabilities. An ideal control method corresponds to an identity matrix (Fig.~\ref{fig:polana}(c)). This produces a click with unity probability at the target detector while keeping the other three detectors in the same row silent. If we just send a very bright pulse below $P_{\rm blind}$ for all the detectors, we will produce a matrix filled with 100\% for all values. This is simply because all detectors are in the linear detection regime and will click with high probability under illumination below $P_{\rm blind}$.

The exact power and polarization of light sent from the setup should be finely adjusted, in order to meet the following requirements: to hit the correct detector, to launch a click with (ideally) 100\% efficiency, and not introduce any double-clicks, such as firing an unwanted detector. Subsequently, the power has to be higher than a certain threshold to launch a click in a blinded detector. For low incident powers, the success rate at the target detector will be lower than unity. This might be interpreted as a virtual optical loss of the connection. A lower success rate can always happen and is  usually compensated by the QKD setup under attack. If the incident power is too high, an unwanted detector might be launched simultaneously as the targeted one, which results in a double click at the receiver. For security reasons in QKD, this has to be treated by replacing the double click with a random bit value~\cite{lutkenhaus1999,tsurumaru2008}. Thereby, an unknown bitflip in the resulting key string is introduced. Subsequently, the knowledge of the eavesdropper becomes less than perfect, hampering her ability to decode the final key. To reliably realize the attack, the introduction of double clicks must be strictly avoided.

The faked-state attack can be implemented, if the response matrix is sufficiently close to an identity matrix. The eavesdropper (Eve) analyzes the single photons sent by the legitimate sender (Alice) and sends a tailored light pulse onwards to the legitimate receiver (Bob) (see Fig.~\ref{fig:polana}(d)). Since Eve and Bob share the same detection events, Eve has the full knowledge of the secret key, when deducing valid events from public channel information.

The scheme described above allows the influence very general quantum-optical experiments. This setup may not only be used just to attack QKD schemes, but also addresses several tests of Bell's inequality~\cite{gerhardt_prl_2011}.

We set the requirements for the device to build as follows, to be universal in the choice of the attack: The blinding power has to be controllable to blind all detectors at desired levels. Further, by applying light pulses both before and after the desired click time, it is possible to achieve a significant degree of control for each detector (as will be explained in Secti n~\ref{sec:control-diagrams}). The implementation described below requires 9 lasers in total: one for the blinding power, keeping all detectors silent, and four times two lasers to target each detector. The device has to be fully programmable and also fast in terms of propagation delay and jitter. 

\section{Device implementation}

\subsection{Optics design}

\begin{figure}
  \includegraphics{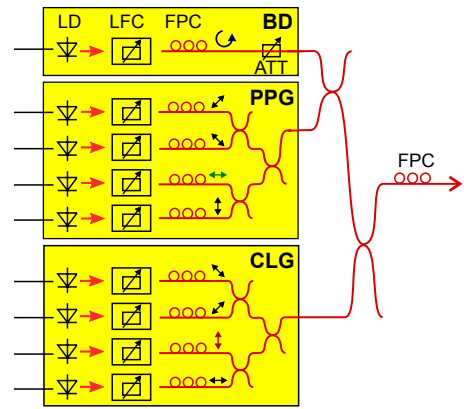}
  \caption {(Color online) Optics part of the experimental setup. Shown are the three relevant parts, the blinding diode (BD), the pre-pulse generator (PPG), and the click-launch generator (CLG). All devices consist of laser diodes (LD). These are combined with laser-to-fiber couplers (LFC) with built-in tunable attenuator. A fiber polarization controller (FPC) allows for control of each polarization. The fiber polarization controller at the output allows for targeting an arbitrary aligned detector of the legitimate receiver in a QKD setup. Each of the fiber couplers is equipped with a manually-adjustable variable attenuator, and the blinding diode is additionally equipped with a programmable variable attenuator (ATT).}
  \label{fig:optics}
\end{figure}

The optical design of the control apparatus is shown in Fig.~\ref{fig:optics}. It is entirely based on laser diodes as light sources. The apparatus was built with fiber-based optics to resist disturbances from the environment and drifts for long-time data acquisitions. This is required for long-time experiments. Each laser diode only produces one optical pulse per control event, and the relative power level for the diodes is not changed. The diodes are either \emph{on} or \emph{off}, driven directly by electronic pulses. This allows for a high driving speed and for a simple electronic driving circuit. The details of the electronic driver are discussed later. The coupling of the diodes is adapted, or variable optical attenuators are used, to change the required power levels. All pulses are combined by fiber beamsplitters to form a control pattern. All laser diodes are single transverse mode diodes and the output light of the setup can be a mixture of differently polarized pulses.

For a certain QKD system~\cite{marcikic_apl_2006,ling_pra_2008,peloso_njp_2009}, we chose the most powerful available single-mode laser diode (Sanyo DL-8141-002) at the wavelength of 808~nm. The light produced by the laser diodes is coupled into single-mode fibers by means of a compact diode-to-fiber coupler (OZ Optics), equipped with a built-in variable attenuator. The light passes polarization controllers (OZ Optics) and is combined by means of fiber beam-splitters. The fiber polarization controllers allow for changing the relative polarization states. For a speficic experimental configuration, a fiber polarization controller at the optical output allows to target an arbitrarily aligned detector.

The coupling efficiency from the diode to fiber reaches up to 60\%. Considering the losses through the fiber path (insertion loss, connector loss and coupling loss of fiber beamsplitters), the overall efficiency throughout the system is about 6\%. The laser diodes have a nominal output power of 200~mW. A maximal power of 12~mW can be achieved at the fiber output of the setup. 
This is orders of magnitude higher than the specified detection range of the single-photon detectors, and much higher than the required blinding power for the single-photon detectors. This allows to compensate for losses in the transmission line to the target detector. Higher output peak-powers can be reached by reducing the number of laser diodes and fiber beamsplitters in the setup.

The overall optical design is based on the combination of nine laser diodes. A single diode, which delivers circularly polarized light, can be used to blind all detectors simultaneously. For convenience, this diode was equipped with a digital variable attenuator (OZ Optics DA-100). This allows for fast change of the power level and can be used for alignment. Four diodes deliver linear polarization from the pre-pulse generator (PPG), which allows for an increased blinding power targeted to any three out of the four detectors. The click-launch generator (CLG) is intended to launch the clicks in the targeted detector. It is formed by another set of four laser diodes, which deliver linear polarization. This allows for maximum flexibility of the control pattern to target the detectors (see Section~\ref{sec:control-diagrams}).

\begin{figure}
  \includegraphics[width=8cm]{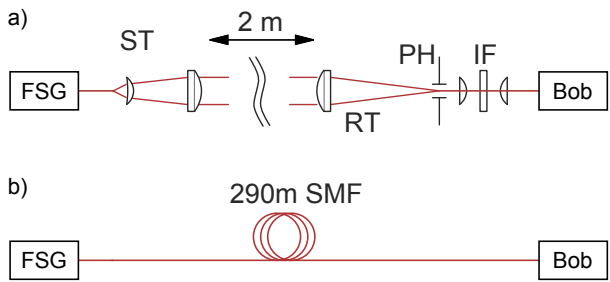}
  \caption{(Color online) The described faked-state generator can be used in a variety of schemes to control single-photon detectors. (a) We set up a free-space experiment with a sending telescope (ST) and receiving telescope (RT). We limited the acceptance angle by an optical pinhole (PH) and acceptance wavelength band by a $10\,\nano\meter$ interference filter (IF). (b) For long-distance experiments, the system was fiber-coupled with single-mode fiber (SMF).}
  \label{fig:experiment}
\end{figure}

Initially, the experiment was carried out in the lab with $2\,\meter$ free-space separation between Eve and Bob (Fig.~\ref{fig:experiment}(a)). It was possible to influence the targeted single-photon detectors in the desired way. However, for long-distance experiments, the coupling efficiency of light from Eve to Bob would be lower and we would need more powerful laser diodes, which were not easily available at the time of our study.

The main experiment was carried out via a $290\,\meter$ long single-mode fiber (see Fig.~\ref{fig:experiment}(b))~\cite{gerhardt_nc_2011}. Before the experiment started, the control diagram shown in Fig.~\ref{fig:ctrldiag}(c) was obtained by a precise tuning. The experiment lasted for about $12\,\hour$ overnight. The alignment had been maintained and no significant drift was observed.

\subsection{Electronics design}

\begin{figure}
  \includegraphics{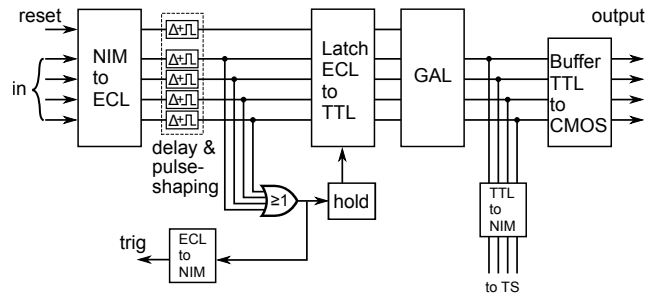}
  \caption {Control circuit for one PPG or CLG. The design is partly derived from Ref.~\onlinecite{gaertnera_rosi_2005}. TS, Eve's time-stamp unit. See text for details.}
  \label{fig:circuitry}
\end{figure}

The electronics was designed to form faked states that have certain sequence of light powers and polarizations, which can be eventually field-adjustable. We call this sequence a \emph{control diagram.} The optical pulses are formed in response to the signal received by a copy of the legitimate receiver unit (Bob$'$, see Fig.~\ref{fig:polana}(d)). The electronic circuit was built to allow for different control diagrams (different pre- and click-launch pulses) and for precise timing. The latter is required since many QKD protocols rely on a precise timing between the legitimate sender and the legitimate receiver, which should be preserved in the presence of an eavesdropper. The electronics has to be able to compensate for different detector time delays and the internal optical propagation delay in the optics part. Further, the electronic circuit has to trigger the internal time-stamping of the eavesdropper, which is done by recording the click-launch pulses sent onwards to Bob. Depending on the exact QKD protocol, it might occur that the eavesdropper receives more clicks than can be sent onwards. 

Two similar electronic circuits were built for the PPG and the CLG. For highest possible flexibility, these included a programmable logic element. An external delay generator (Highland Technology P400) was used for different longer time-delays, such as setting the laser pulse lengths.

Each of these custom-built circuits (see Fig.~\ref{fig:circuitry}) consists of 4 independent input channels and a dedicated reset line. The input signals from the detector have nuclear instrumentation (NIM) logic levels, $0$ and $-16\,\milli\ampere$ into a $50\,\ohm$ load. At the input, they are converted to emitter-coupled logic (ECL) levels, $-0.9$ and $-1.75\,\volt$ by a transistor (BFR93) and a differential receiver (MC100EL16). The resulting ECL signals are combined on a logic OR element (MC100EL01) and launch a trigger pulse {\it trig} for an external delay generator. Tunable delay stages (trimmer-adjustable one-shot triggers) are used to compensate different arrival timing of pulses from the different channels. The ECL signals are converted into transistor-transistor logic (TTL) levels by a converter (MC100H603), and processed by a generic array logic (GAL) integrated circuit (GAL16V8). The GAL is programmed to produce a rising edge on the desired channel and to produce a falling edge by the reset line, triggered by the external delay generator. The function of the circuit can be easily modified by using differently programmed GALs. The external delay generator defines the output pulse width with sub-nanosecond precision by activating the {\it reset} line. The pulses produced by the GAL are buffered and amplified with a complementary metal-oxide-semiconductor (CMOS) line driver (74ACT11004), which directly drives the laser diodes via small current-limiting resistors. The design for this control circuit was partly derived from Ref.~\onlinecite{gaertnera_rosi_2005}.

\begin{figure}
  \includegraphics{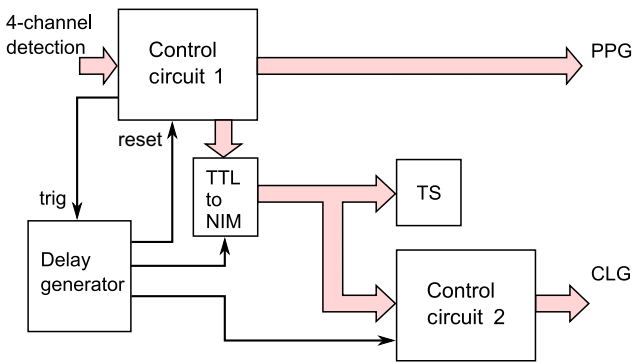}
  \caption {(Color online) Electronics part of the experimental setup. Eve receives the photon and decodes it into one of the four polarization states. The electrical pulse (``click'') is processed by a control circuit~1, to equalize internal delays in the system and to control the (optional) pre-pulse generator (PPG). The output of the control circuit~1 triggers a delay generator followed by control circuit~2 for the click-launch generator (CLG), and also sends the clicks onwards to Eve's time-stamp unit TS.}
  \label{fig:electronics}
\end{figure}

In our experiment, one of the circuits described above takes the input from the receiver unit Bob$'$, and its output drives the laser diodes in the PPG (Fig.~\ref{fig:electronics}). The second circuit is cascaded to the first one and receives its input from the GAL in the first circuit. This circuit drives the CLG. The delay generator is triggered by the first circuit and sets both the pulse width of the PPG and the pulse delay and width of the CLG. For recording the actual faked states, the time-stamp unit TS of the eavesdropper is attached to the output of the first circuit. This ensures that only events that are actually sent onwards are recorded. Since the optical design of the PPG and the CLG is equal, it was possible to run all control diagrams discussed in the next section with the above described optical and electronic configuration. Just the GAL was reprogrammed. The PPG was used as the only source to generate pulses, for simple control diagrams with only one optical pulse per received click, such that it was used as a CLG.

\section{Results}
\subsection{\label{sec:control-diagrams}Control diagrams} 

\begin{figure*}
  \includegraphics{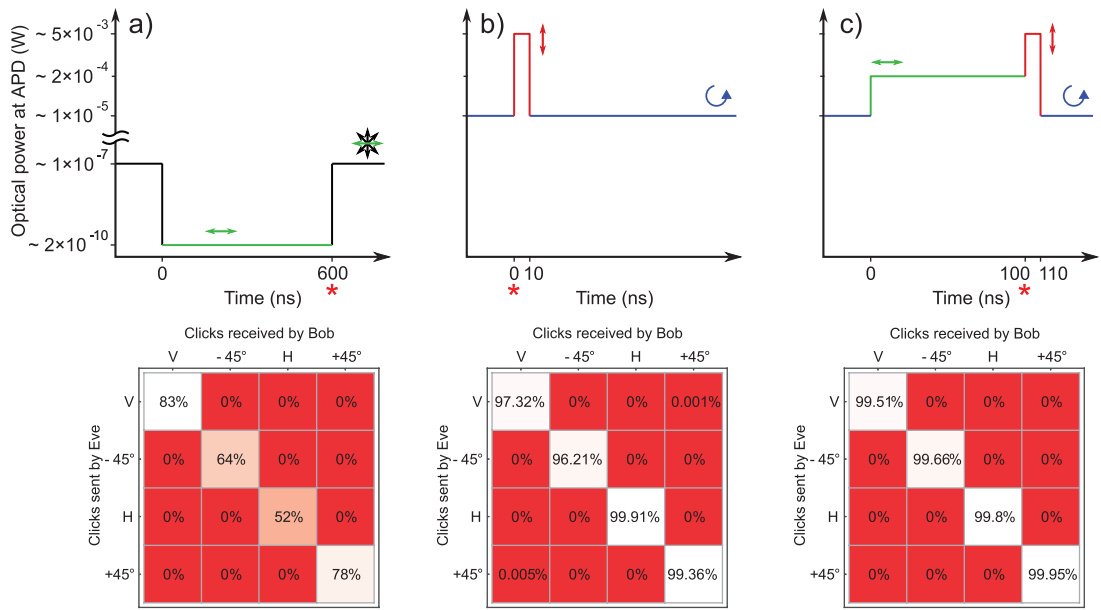}
  \caption{(Color online) Tested control diagrams. The plots show Eve's output optical power and polarization as a function of time. Timing of Eve's received click is denoted as $0$. The timing of click in the target detector is marked with asterisk. The matrices show the control fidelity for the three control diagrams. The experimental complexity is increased from left to right: (a)Diagram 1: 4 lasers selectively switched off; (b)Diagram 2: 5 lasers: 1 for blinding and 4 for click-launching pulses; (c)Diagram 3: 9 lasers: previous plus 4 extra leasers for forming pre-pulses.}
  \label{fig:ctrldiag}
\end{figure*}

The constructed device allowed for different control schemes of single-photon detectors. Initially, it was unclear if an unwanted cross-talk between multiple detectors on different polarizations leads to accidental clicks. For simplicity, some of our experiments were conducted with simple protocols (such as described in Ref.~\cite{gerhardt_prl_2011}). Further refinements of the control diagram allowed to reach higher efficiencies and to reduce the number of accidental clicks in the case of a slight misalignment. These more robust protocols were used in our experiment, in which we conduct an attack on QKD that was carried out partially outdoors~\cite{gerhardt_nc_2011}. It led to eavesdropping the entire secret key. In total, three control diagrams have been characterized. The protocols are described below.

\paragraph{Diagram~1:} The first control diagram was proposed earlier~\cite{makarov_njp_2009}. All four detectors are blinded by constant illumination of mixed linear polarizations from 4 lasers. A click at the target detector is realized by introducing a temporal gap. The polarization of illumination corresponds to the orthogonal-basis detector (only the corresponding laser is kept on, while the other three are switched off), see Fig.~\ref{fig:ctrldiag}(a). The power is adjusted to a level where the detectors in the conjugate basis are kept blinded. At the end of the gap, only the target detector recovers sensitivity and clicks when the power is restored, while the other three remain silent. The click probability at the target detector with a gap of 600~ns is only 52--83\% (Fig.~\ref{fig:ctrldiag}(a)). This lack of efficiency is not a problem, since the reduced efficiency would be interpreted as an additional optical loss in the QKD experiment and would be simply reducing the bitrate. This control diagram was employed in the study of non-physical Bell tests~\cite{gerhardt_prl_2011}. Since this diagram is targeting passively-quenched APDs, the timing relates to the recovery time of the APD response. This time is approximately $1\,\micro\second$. The click efficiency is forming a sigmoidal curve, which starts at zero for gap-times below 300--400~ns, and which reaches up to unity, when the gap-time is increased above microseconds. Minimally four laser diodes and one of the aforementioned circuits are required for this protocol. A more powerful control diagram is proposed below, attempting to overcome the two disadvantages of diagram~1: low click probability and long gap-time. 

\paragraph{Diagram~2:} All four detectors are blinded by constant illumination with circular polarization. A click is launched at the target detector by a very bright (mW range) polarized pulse instead of a timing gap (see Fig.~\ref{fig:ctrldiag}(b)). The power of the pulse is adjusted to not launch the other three detectors. Some unwanted clicks still existed in practice, as illustrated by the small off-diagonal elements in the matrix. The main reason is that the four detectors do not necessarily have identical optical and electrical characteristics. This diagram is significantly more compact in time than diagram~1. The requirements for this diagram are five laser diodes and one control circuit.

\paragraph{Diagram~3:} To eliminate unwanted clicks, a pre-pulse polarized orthogonally in respect to the target detector is introduced $100\,\nano\second$ before the click-launch pulse (Fig.~\ref{fig:ctrldiag}(c)). This blinds the non-target detectors deeper and makes it more difficult to be launched by the main pulse. The deeper blinding occurs via reducing the voltage across the APD much lower than strictly necessary for blinding. At low voltages, the APD has smaller finite multiplication gain than when biased just below its breakdown voltage. The off-diagonal elements in the matrix are eliminated. Although the click probability at the target detector is slightly lower than 100\%, the eavesdropper can still get full information on the key by listening to the key sifting procedure during classical communication part of the QKD protocol. This control diagram is the final implementation to our eavesdropper setup. For this diagram, nine laser diodes and the full electronic scheme as shown in Fig.~\ref{fig:electronics} are required.

Further control diagrams are feasible with the presented setup. All diagrams rely on either switching off a laser for a given detector or increasing the power significantly. The remaining detectors see less laser power than the targeted one. To allow for an even higher efficiency in unfavorable conditions, it might be possible to increase the blinding power further before an actual click-launch event. This might be analyzed in further studies.

\subsection{Alignment procedure}

In order to achieve good control of Bob's detectors, polarizations and powers of Eve's lasers have to be individually adjusted for each particular receiving unit. The alignment procedure is carried out as three steps: First, the circular blinding power is adjusted. Then, the pre- and later the click-launch pulses are aligned. To emulate a copy of the detector unit (Bob$'$), a simple pulse generator at a frequency of 10~kHz was used, producing 15~ns wide pulses similar to detector output pulses. There is a waiting time of $100\,\micro\second$ between consecutive pulses, which is a relaxed condition comparing to what a real Bob$'$ can output (as further discussed below).

\begin{figure*}
  \includegraphics[width=0.8\textwidth]{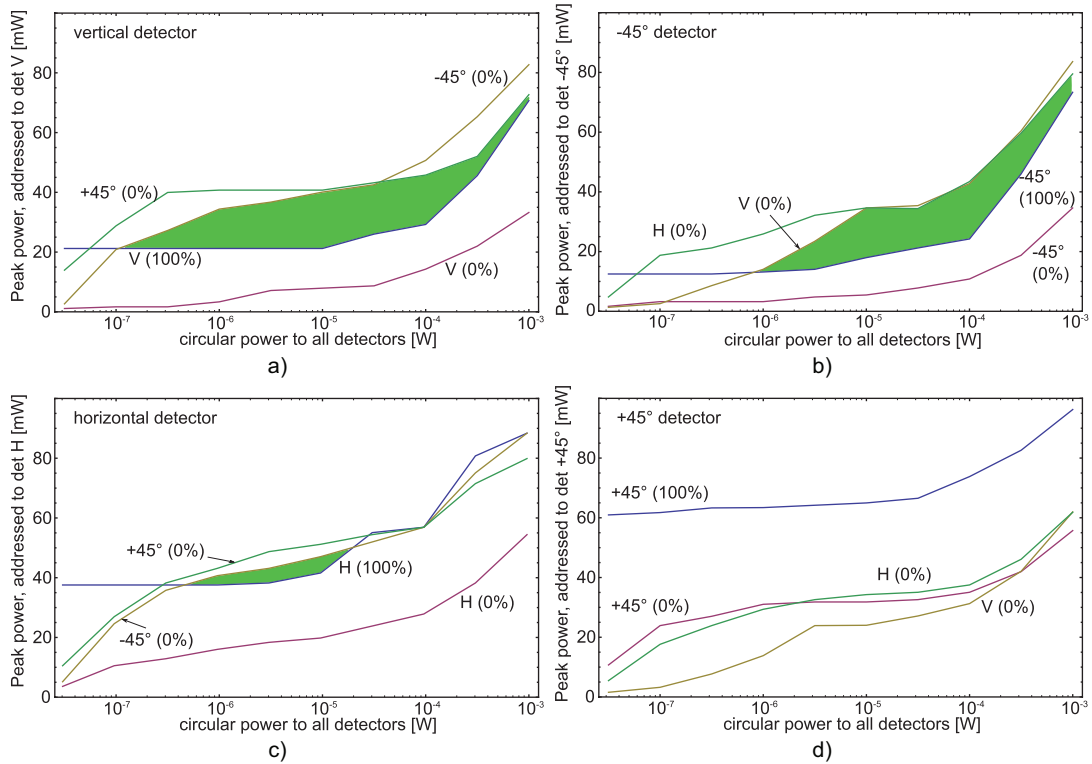}
  \caption{(Color online) Thresholds of silence and 100\% click probability for diagram~2. For alignment purposes the exact background levels and peak-powers have to be targeted to the specific detectors. The graphs show the click probability of the targeted detector (both for 0 and 100\% click probability) and the non-orthogonal detectors, which are also sensitive to the incoming laser. The aim is to launch a click in the targeted detector with unity efficiency, while keeping all other detectors in the blind state.}
  \label{fig:perfoc}
\end{figure*}

The goal is to implement the aforementioned diagram~3. For alignment of the circularly polarized blinding power, the blinding diode is turned on in continuous-wave mode. It is ensured that all APDs are blinded by adjusting the electronic controllable attenuation. The alignment is complicated by the unequal blinding characteristics of Bob's APDs. A better control is obtained by distributing blinding light unevenly before them. This can be achieved by a slightly elliptical polarization. The polarization is adjusted by monitoring Bob's count rates in the linear, unsaturated regime, below $P_{\rm blind}$. Only the targeted detector should be launched at 100\% efficiency. This task could not be achieved with only one circularly (or slightly elliptically) polarized blinding light. Ideally, the blinding power is set to a reasonable level, such that the unbalance of four APDs can be compensated most. Then, a unity matrix can be achieved after introducing pre-pulses. We introduce the alignment for diagram~2 described in the following paragraphs. The blinding power is found based on measurements of diagram~2, as shown in Fig.~\ref{fig:perfoc}.

\begin{figure*}
  \includegraphics[width=112mm]{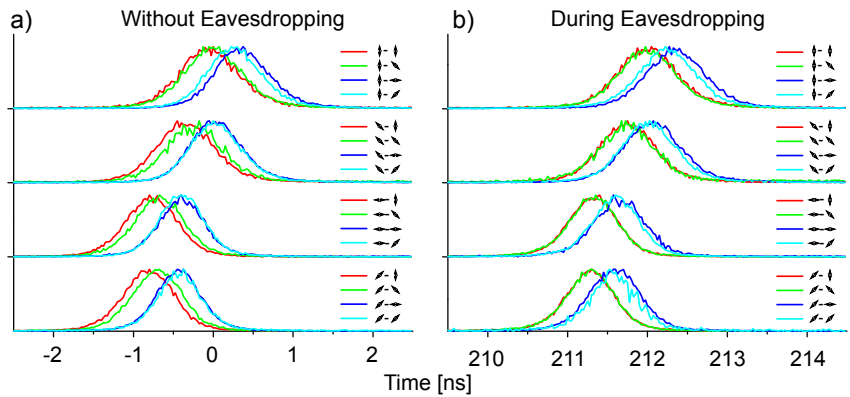}
  \caption{(Color online) Histogram of time delays between all coincidence polarization combinations, as measured by Alice and Bob without Eve (a) and with Eve inserted into the line (b). The full width at half maximum (FWHM) averaged over the 16 peaks is $761\,\pico\second$ in the former case and $779\,\pico\second$ in the latter~\cite{gerhardt_nc_2011}.}
  \label{fig:delay}
\end{figure*}

It is instead required to align the click-launch pulses right after setting the blinding power for the simpler structured diagram~2. In Fig.~\ref{fig:perfoc}, the blinding power is displayed against the peak-power of the click-launch pulse for each of the three detectors which receive a part of the incident power (see Fig.~\ref{fig:polana}). The fourth detector does not receive any light and will remain silent throughout the experiment. For each of the three illuminated detectors the threshold of silence (labeled as 0\%) is displayed. For the targeted detector, we also introduced a curve, which shows when the detector is fired with unity probability (labeled as 100\%). The alignment has to ensure, that only the targeted one is launched with 100\% efficiency, while the others remain silent. We have to align both the circular blinding power and the peak-power. For both settings, there is an optimal point. The windows between the power for reaching unity click efficienty and silent other detectors should be large. This equals a blinding power of around 10$^{-5}$~W for the displayed situations. See Fig.~\ref{fig:perfoc}(a) as an example. This ensures the stability of the apparatus, also for slight changes in the optical alignment. In Fig.~\ref{fig:perfoc}(d), the main problem of diagram~2 becomes evident: there is no such window. This is likely due to variation in individual APD's characteristics. Due to the lack of such a window, it is impossible to reach an optimal alignment with unity click-efficiency for the targeted detector. If we increase the click efficiency, we will introduce cross-talk to other detectors. This can be avoided by using pre-pulses.

After setting the blinding power, it is required to set the pre-pulse level. These pre-pulses are used to decrease the sensitivity against further illumination. It is important that the pre-pulses do not launch clicks on their own. During adjustment, the power of pre-pulses is increased until they start launching clicks in some of the detectors. Subsequently, their level is slightly reduced.

To generate a click in diagram~3, it is required to have additional pulses to actually launch a click in the targeted detector. These pulses were generally very powerful (tens of mW) and the pulse powers were adjusted as a final alignment step. For a 10~kHz excitation rate with a pulse generator, the power is increased until clicks were received on the targeted detector. The power is further increased to reach an efficiency of unity. In most cases with diagram~3 this is sufficient to reach a diagonal matrix with unity click efficiency, since the power level of blinding light and pre-pulses have already been adjusted. A cross-talk between adjacent pulses led to slightly less than 100\% click probability in the measured matrix.

The alignment was accomplished by inserting Eve into the fiber line before attack and manually calibrating her polarizations and power levels to match Bob's detector settings. The calibration only requires that the transmission on the classical channel is public, which is one of the basic assumptions of QKD. The stability has been proven by the fact that the alignment kept steady for $\sim$12 hours in an overnight experiment. To be more robust, this alignment procedure could be automated and implemented in a non-obtrusive way when the link is constantly running~\cite{makarov2005}.

\subsection{Timing analysis}

The introduction of an eavesdropper into a quantum cryptographic connection will also affect the timing between the two communicating parties. It is required that Alice and Bob have a common clock. This is commonly realized by an external timing reference e.g. from the global positioning system, or by atomic clocks. In the system under attack, the timing is servoed over the joint detection events. These originate from the accurate timing of the entangled photons used for QKD protocol. It is not required to have an atomic time reference for initial negotiation~\cite{ho_njp_2009}. In the following we will discuss timing issues between the communication partners and the disturbance introduced by an eavesdropper.

Insertion delay of the eavesdropper is an important change in the communication protocol. For our system, we measured insertion delay of about $212\,\nano\second$ (see Fig.~\ref{fig:delay}). Its main sources are the pulse length of the required laser pre-pulses and the electronic and optical propagation delay in the system. It is possible to compensate for this insertion delay. As a lower limit it is evident that the detection event of Bob$'$ has to be long enough to trigger an electronic circuit to drive the laser diodes. An additional delay will be introduced by the population inversion build-up time of the semiconductor laser diode. For the laser pulses, the circuit described in this paper is designed to fully turn the diodes on and off. It will be faster to drive the diode from just below the lasing threshold to above the lasing threshold, as commonly done in telecommunication transmitters. We estimate a lower limit for a delay of the eavesdropper to be within $10$--$30\,\nano\second$. This assumes normal off-the-shelf available electronic components and laser diodes. It would be possible to shortcut a part of the communication line with free-space line-of-sight radio-frequency or an optical link for fiber based system~\cite{makarov2005}. The propagation delay in such link is $n \approx 1.5$ times shorter than in optical fiber of equal length. This would not help in case of a free-space QKD system.

In an ideal case, all four photodiodes in the legitimate detector unit have the same optical and electrical delay, however in practice the delays are not the same and have to be compensated in software post-processing~\cite{lamas-linares_oe_2007}. Changes in these relative delays between different detector combinations Alice--Bob are another possible signature of Eve's presence. By careful adjustment of the tunable delay stages in Eve's electronics, these relative delays were kept unchanged (as evident in Fig.~\ref{fig:delay}).

\begin{figure}
  \includegraphics[width=\columnwidth]{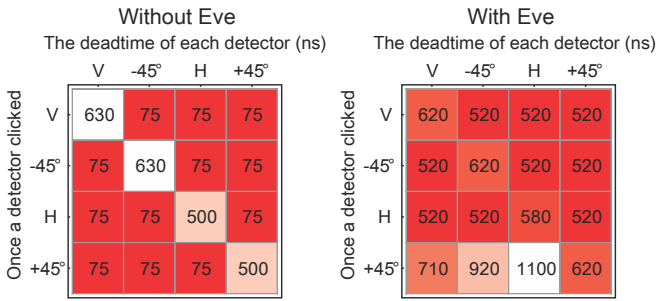}
  \caption{(Color online) Deadtime of all 16 detector combinations (auto-correlation and cross-correlation) with and without eavesdropping.}
  \label{fig:deadtime}
\end{figure}

Each detector has a distinct deadtime after click, during which it is not single-photon sensitive, since the voltage level is below the detection threshold and the pulse-discriminator has just fired. Besides this, the deadtime might also be influenced by other detector channels, because of limitations of Bob's click registration system (cross-channel effect). The introduction of an eavesdropper has varied these deadtimes (see Fig.~\ref{fig:deadtime}). For a detector unit, the pulses \emph{for different detectors} have a very short time-delay of $\lesssim 100\,\nano\second$. This is the lower limit, given by the deadtime of the time-stamp unit used for these experiments. We have artificially introduced $520\,\nano\second$ deadtime, to reduce uneven responsivity of detectors recovering from the previous launch- and pre-pulses. For the \emph{same detector}, the deadtime is $\sim 1\,\micro\second$, simply introduced by the recovery of the APDs responsivity to the trigger pulse.

\section{Detectability of QKD intrusion}

Since the apparatus is targeting QKD systems, the question arises if such an intrusion event can be detected by Bob, or even by Alice in the QKD communication. The demonstrated attack is well-documented and several countermeasures are evident. It might be possible to obscure all countermeasures by smarter attack design. A variety of attacks into QKD connections exist already. By \emph{not} detecting the below mentioned side-effects it \emph{cannot} be guaranteed that the connection is secure.

Since we introduce a bright-light source to blind the detector efficiently, it will be possible to detect the blinding light directly with an additional photodiode in the polarization detector and only allow for single-photon counting if the background light level allows for single-photon counting. Also, the current drawn by the APD power supply will be increased, since the diode voltage level is constantly reduced, such that the system would draw more current. In this sense, a simple current monitor might be sufficient to detect this intrusion.

An eavesdropper with own pulse-shaping (a few hundred ns) and electrical delay (a few hundred ns) will introduce a few hundred ns delay between Alice and Bob. Depending on the exact configuration, it can be possible to measure, e.g., the distance between the legitimate communicating parties, and allow only for a tight ranges of transfer delay. An eavesdropper would exceed this bound and the connection would be regarded as insecure. In the QKD system we attacked, the exact time synchronization is servoed over the detection events. Only for initial synchronization two external referenced clocks were used. By carefully introducing longer and longer delays, it might be possible to keep the connection up and running and introduce the eavesdropper. 

Relative timing between Alice and Bob has a certain jitter. An eavesdropper should seek not to change the present jitter too much. Eve's FSG and the response of the legitimate detector unit to Eve's control constitute an additional source of jitter under attack. This jitter may or may not be smaller than detector's intrinsic jitter in the single-photon detection regime. In the case of the QKD system we tested, Bob's jitter under control was much lower, down to $\sim 50\,\pico\second$ full width at half maximum (FWHM), than its intrinsic single-photon detection jitter of $\sim 500\,\pico\second$ FWHM. (We remark that the reduction of jitter under control has also been observed in SNSPDs~\cite{tanner2013}.) The system we tested had $761\,\pico\second$ FWHM jitter averaged over all cross-correlations under normal operation without the presence of the eavesdropper, while with the eavesdropper it increased to $779\,\pico\second$ FWHM (Fig.~\ref{fig:delay}). The difference is at the lower limit of detectability. Note that the eavesdropper might compensate the increased jitter by reducing the jitter of her own detectors.

In the above design, it is evident that the legitimate detector unit (Bob) will not receive any double clicks on different detectors. This is also the case for clicks that have a very short time delay between different polarization bases. Only one pulse will be given out per detection event and the pulse length ($\approx 110\,\nano\second$ in case of diagram~3) is an effective deadtime. If the QKD system used a time-stamp unit with no or reduced inter-channel deadtime, it would be possible to observe a modified photon statistics in the presence of an eavesdropper. This holds especially for entanglement-based QKD schemes, but is not an issue with prepare and send schemes such as BB84 with a low clock rate. In further versions of this apparatus, it will be possible to trigger any photon pattern on a target detector. This requires more laser diodes and a more sophisticated electronic.

The optical fine-tuning is conducted such that we reach a diagonal blinding matrix with unity click efficiency. If the eavesdropper is very close to the sending unit, the optical loss between them might be very low. This would result in a detection rate of Bob$'$ exceeding Bob's usual rate. In the actual experiment, it is required to reduce the coupling efficiency on the incoming side of the eavesdropper, such that the bitrate is matching with the situation without an eavesdropper.

Similarly, there is a problem, if there is a slight mismatch of detector efficiencies at the legitimate receiver. In the attack scenario of an eavesdropper generating faked states, any mismatch is hidden by the unity click efficiency. Subsequently, any mismatch in eavesdropper's copy of the legitimate receiver unit is replicated onto Bob's detectors. This might introduce statistical fingerprints, which might be evident to Alice and Bob.

The intermediate position of an eavesdropper might also be deducted when the APD detection flashback is analyzed. An ideal photodiode would not emit any light when it is fired, but in realistic devices, APDs emit a small flash, when receiving a photon and launching a click~\cite{kurtsiefer_jmo_2001}. If the receiving unit of the eavesdropper is closer than the legitimate receiver unit, Alice might receive a flashback from Eve's detection events. This event will be earlier detectable than with the legitimate detector unit of Bob.

By the countermeasures mentioned above, it is relatively simple to detect eavesdropping by different measures. In many implementations of QKD the countermeasures discussed are not implemented and as a consequence they may be vulnerable to the types of attack demonstrated in the paper. And even if all the above countermeasures are introduced, there might be other ways of intruding into such systems. An alternative approach to solving detector vulnerability problems is to employ device-independent heralded qubit amplifier~\cite{gisin2010, kocsis2013} or measurement-device-independent~\cite{lo2012,rubenok2012} QKD protocols. Once these have been fully implemented, our intrusion will face strong challenges, since the security will no longer rely on properties of single-photon detectors.

\section{Conclusion}

In this paper we have described a technical apparatus for control of various single-photon polarization analyzers, not only attacking QKD schemes, but also influencing very general quantum-optical experiments, e.g., tests of Bell's inequality. It was successfully used to control the outcome of passively-quenched single-photon detectors in several experiments~\cite{gerhardt_nc_2011, gerhardt_prl_2011}. For polarization-encoded qubits, such as in QKD, it was possible to target a single detector while keeping the other detectors silent. The design goal was to optimize the click probability to unity and to reduce the cross-talk to the other detectors to zero. We further discussed how an optimal alignment can be reached, and possible side effects of the attack.

In future studies, this apparatus will allow to study the saturation behavior and linearity of various photodetectors, test countermeasures to detector control attacks and hacking-resistant QKD schemes. The photodetector active control system developed provides power outputs of up to $\sim 100\,\milli\watt$ for each receiver element. This provides sufficient flexibility for testing photodetectors with a wide range of response characteristics.

\begin{acknowledgments}
This work was supported by the National Research Foundation and the Ministry of Education, Singapore, and the Research Council of Norway (Grant No.\ 180439/v30). V.M.\ thanks Industry Canada for later support. I.G.\ thanks J.\ Wrachtrup for continuous support.
\end{acknowledgments}

\end{document}